\documentclass[aps,prd,amsmath,amssymb,preprintnumbers,nofootinbib,a4paper,preprint]{revtex4-1}
\pdfoutput=1
\usepackage{amsthm}
\usepackage{graphicx}
\usepackage{color}
\newcommand{\beq}{\begin{eqnarray}}
\newcommand{\eeq}{\end{eqnarray}}

\newcommand{\bmp}{\noindent\begin{minipage}{16cm}}
\newcommand{\emp}{\end{minipage}\vskip 7mm} 

\usepackage{dcolumn}
\usepackage{bm}
\usepackage{bbm}
\usepackage{subfigure}
\usepackage{pxfonts}

\theoremstyle{definition}

\theoremstyle{plain}

\usepackage{epsfig}

\usepackage{graphicx}
\usepackage{multirow}
\usepackage[margin=5pt, font=normalsize,labelfont=bf,format=plain,justification=centerlast]{caption}

\definecolor{rossoCP3}{cmyk}{0,.88,.77,.40}

\def\lsim{\mathrel{\rlap{\lower4pt\hbox{\hskip1pt$\sim$}}
    \raise1pt\hbox{$<$}}}                
\def\gsim{\mathrel{\rlap{\lower4pt\hbox{\hskip1pt$\sim$}}
    \raise1pt\hbox{$>$}}}                

\newcommand{\be}{\begin{eqnarray}}
\newcommand{\ee}{\end{eqnarray}}

\usepackage[
bookmarks]{hyperref}
\usepackage{color}
\hypersetup{colorlinks, bookmarksopen, bookmarksnumbered,
citecolor=verdes, linkcolor=bluc, pdfstartview=FitH, urlcolor=rossos}

\definecolor{grigio}{cmyk}{0,0,0,0.1}
\definecolor{rosa}{cmyk}{0,0.1,0.1,0.02}
\definecolor{rosino}{cmyk}{0,0.05,0.05,0.02}
\definecolor{rosas}{cmyk}{0,0.3,0.25,0.05}
\definecolor{celeste}{cmyk}{0.1,0,0,0.02}
\definecolor{giallino}{cmyk}{0,0,0.1,0.02}
\definecolor{rosso}{cmyk}{0,1,1,0.4}
\definecolor{rossos}{cmyk}{0,1,1,0.55}
\definecolor{rossoc}{cmyk}{0,1,1,0.2}
\definecolor{blu}{cmyk}{1,1,0,0.3}
\definecolor{blus}{cmyk}{1,1,0,0.5}
\definecolor{bluc}{cmyk}{1,1,0,0.1}
\definecolor{blucc}{cmyk}{0.7,0.5,0,0}
\definecolor{viola}{cmyk}{0,1,0,0.6}
\definecolor{viola2}{cmyk}{0,1,0.2,0.6}
\definecolor{verde}{cmyk}{0.92,0,0.59,0.25}
\definecolor{verdec}{cmyk}{0.92,0,0.59,0.15}
\definecolor{verdes}{cmyk}{0.92,0,0.59,0.4}
\definecolor{verdino}{cmyk}{0.12,0,0.09,0.02}
\definecolor{giallo}{cmyk}{0,0,1,0}
\definecolor{gialloverde}{cmyk}{0.44,0,0.74,0}
\definecolor{Titolo}{rgb}{0.752941176,0.576470588,0.992156863}
\definecolor{altro}{rgb}{0.094117647,0.650980392,0.643137255}
\definecolor{Peanuts}{rgb}{0.2, 0.4, 0.6}
\definecolor{Pean1}{rgb}{0.6, 0.8, 0.4}
\definecolor{BHO}{rgb}{0.2, 0.8, 1}
\definecolor{Daria}{rgb}{0, 0.9412, 0}
\definecolor{UniPi}{rgb}{0.2549, 0.4627, 0.6275}
\definecolor{UniPidue}{rgb}{0.3216, 0.5804, 0.7882}

\DeclareMathOperator{\Tr}{Tr}

\usepackage{youngtab}
\usepackage{slashed}

\definecolor{rossoCP3}{cmyk}{0,.88,.77,.40}

\begin{document}
\title{\Large  \color{rossoCP3} Fundamental Composite Higgs Dynamics on the Lattice:\\ SU(2) with Two Flavors}
\author{Ari Hietanen$^{\color{rossoCP3}{\varheartsuit}}$}\email{hietanen@cp3-origins.net} 
\author{Randy Lewis$^{\color{rossoCP3}{\spadesuit}}$}\email{randy.lewis@yorku.ca}
\author{Claudio Pica$^{\color{rossoCP3}{\varheartsuit}}$}\email{pica@cp3-origins.net} 
\author{Francesco Sannino$^{\color{rossoCP3}{\varheartsuit}}$}\email{sannino@cp3-origins.net} 

\affiliation{
\vspace{5mm}
{$^{\color{rossoCP3}{\varheartsuit}}${ \color{rossoCP3}  \rm CP}$^{\color{rossoCP3} \bf 3}${\color{rossoCP3}\rm-Origins} \& the {\color{rossoCP3} \rm Danish IAS},
University of Southern Denmark, Campusvej 55, DK-5230 Odense M, Denmark}
\vspace{5mm} \\
\mbox{ $^{\color{rossoCP3}{\spadesuit}}$Department of Physics and Astronomy,  {\color{rossoCP3} York University}, Toronto, M3J 1P3, Canada}
\vspace{1cm}
}

\begin{abstract}
In reference \cite{Cacciapaglia:2014uja} a unified description, both at the effective and fundamental Lagrangian level, of models of composite Higgs dynamics was proposed. In the unified framework the Higgs itself can emerge, depending on the way the electroweak symmetry is embedded, either as a pseudo-Goldstone boson or as a massive excitation of the condensate. 
The most minimal fundamental description consists of an SU(2) gauge theory with two Dirac fermions transforming according to the defining representation of the gauge group. We therefore provide first principle lattice results for the massive spectrum of this theory. We confirm the chiral symmetry breaking phenomenon and determine the lightest spin-one axial and vector masses. The knowledge of the energy scale at which new states will appear at the Large Hadron Collider is of the utmost relevance to guide experimental searches of new physics.
 \\ ~ \\
[.1cm]
{\footnotesize  \it Preprint: CP3-Origins-2014-012 DNRF90  \& DIAS-2014-12}
\end{abstract}

\maketitle

\section{Introduction}

The Standard Model (SM) of particle interactions successfully describes Nature. However, the SM  is unappealing. For example the SM Higgs sector simply models spontaneous symmetry breaking, it does not explain it. Furthermore, there is no consistent way to protect the electroweak scale from higher scales, leading to the SM naturalness problem. We refer to \cite{Antipin:2013exa} for a mathematical classification of different degrees of naturality.  

It is well known that by replacing the SM Higgs sector with a fundamental gauge dynamics featuring fermionic matter fields renders the SM Higgs sector natural. Technicolor \cite{Weinberg:1975gm,Susskind:1978ms} is a time-honored incarnation of this idea. Other ways to use fundamental dynamics to replace the SM Higgs sector appeared later in \cite{Kaplan:1983fs,Kaplan:1983sm}.  The Technicolor Higgs \cite{Sannino:1999qe,Hong:2004td,Dietrich:2005jn,Dietrich:2005wk,Sannino:2009za}
 is the lightest scalar excitation of the fermion condensate responsible for electroweak symmetry breaking. The interplay between the gauge sector and the SM fermion mass sector is relevant because it can reduce the physical mass of the Technicolor Higgs \cite{Foadi:2012bb}. If the underlying dynamics has a larger global symmetry group than the one strictly needed to break the electroweak symmetry successfully, one may be able to choose an electroweak embedding in a way that the electroweak symmetry remains intact. Differently from the Technicolor case, here the Higgs state could be identified with one of the Goldstone Bosons (GB) of the theory. In this case the challenges are not only to provide masses to the SM fermions but also to break the electroweak symmetry by means of yet another sector which can also contribute to give mass to the would--be pseudo-GB Higgs.  Real progress with respect to the SM Higgs sector shortcomings is achieved, however, only if a more fundamental description exists.

In reference \cite{Cacciapaglia:2014uja} a first unified description of models of electroweak composite dynamics was put forward. The description clarified the main similarities, interplay, and shortcomings of the different approaches. In addition a specific underlying realization in terms of fundamental strongly coupled gauge theories was investigated with a clear link to first principle lattice simulations. It was also shown that for a generic electroweak vacuum alignment, the observed Higgs is neither a purely pGB state nor the Technicolor Higgs, but a mixed state. This result has relevant implications for its physical properties and associated phenomenology.

Given a possible underlying gauge theory featuring fermionic matter one can imagine distinct patterns of chiral symmetry breaking \cite{Peskin:1980gc,Preskill:1980mz,Kosower:1984aw,Sannino:2004qp,Dietrich:2006cm,Sannino:2009aw,Mojaza:2012zd}. First principle lattice simulations are now in a position to answer these questions \cite{Catterall:2007yx,Catterall:2008qk,DelDebbio:2008wb,DelDebbio:2008zf,Catterall:2009sb,Hietanen:2009az,DelDebbio:2009fd,Kogut:2010cz,Karavirta:2011zg,Lewis:2011zb,Hietanen:2012qd,Hietanen:2012sz,Hietanen:2013fya,Hietanen:2013gva}.

The classification of underlying gauge theories relevant for Technicolor models appeared in \cite{Dietrich:2006cm}, while for composite models of the Higgs as a pGB can be found in~\cite{Mrazek:2011iu,Bellazzini:2014yua}. In reference~\cite{Cacciapaglia:2014uja} it was concluded that from the point of view of a fundamental theory with fermionic matter, the minimal scenario to investigate is SU(4)$\to$Sp(4) (locally isomorphic to SO(5)), for both a minimal Technicolor as well as composite GB Higgs scenario. The difference being in the way one embeds the electroweak theory within the global flavor symmetry. This pattern of chiral symmetry breaking can be achieved dynamically via an underlying SU($2$)=Sp($2$) gauge theory with 2 Dirac flavors (i.e. four Weyl fermions) transforming according to the fundamental representation of the gauge group. 

We will provide here the state-of-the-art lattice results confirming the breaking of the global SU(4) symmetry to Sp(4) (locally isomorphic to SO(5)), first observed in \cite{Lewis:2011zb}, via the formation of a non-perturbative fermion condensate, and in addition we will further determine the spin-one spectrum.  

The paper is organized as follows: In section~\ref{sec:lattheory} we introduce the lattice framework and detail how the lattice computations of the spectrum is performed; In section~\ref{sec:latresults} the numerical results are summarised; Finally we offer our conclusions in section~\ref{sec:conclusions}. 

\section{The Lattice Method}\label{sec:lattheory}

In the continuum, the Lagrangian for our technicolor template is
\begin{equation}
{\cal L} = -\frac{1}{4}F_{\mu\nu}^aF^{a\mu\nu}
         + \overline{u}(i\gamma^\mu D_\mu-m_u)u
         + \overline{d}(i\gamma^\mu D_\mu-m_d)d
\end{equation}
which can be discretized in the familiar way to arrive at a Wilson action,
\begin{eqnarray}
S_W &=& \frac{\beta}{2}\sum_{x,\mu,\nu}\left(1-\frac{1}{2}{\rm ReTr}U_\mu(x)
        U_\nu(x+\hat\mu)U_\mu^\dagger(x+\hat\nu)U_\nu^\dagger(x)\right)
      + \sum_x\overline{\psi}(x)(4+m_0)\psi(x) \nonumber \\
   && - \frac{1}{2}\sum_{x,\mu}\left(\overline{\psi}(x)(1-\gamma_\mu)U_\mu(x)\psi(x+\hat\mu)
   +\overline{\psi}(x+\hat\mu)(1+\gamma_\mu)U_\mu^\dagger(x)\psi(x)\right) \,, 
\end{eqnarray}
where $U_\mu$ is the gauge field and $\beta$ the gauge coupling in conventional
lattice notation.  $\psi$ is the doublet of $u$ and $d$ fermions, and $m_0$ is
the 2$\times$2 diagonal mass matrix.


Mesons will couple to local operators of the form
\begin{eqnarray}
{\cal O}_{\overline{u}d}^{(\Gamma)}(x) &=& \overline{u}(x)\Gamma d(x) \,, \\
{\cal O}_{\overline{d}u}^{(\Gamma)}(x) &=& \overline{d}(x)\Gamma u(x) \,, \\
{\cal O}_{\overline{u}u\pm\overline{d}d}^{(\Gamma)}(x)
    &=& \frac{1}{\sqrt{2}}\bigg(\overline{u}(x)\Gamma u(x)
        \pm \overline{d}(x)\Gamma d(x)\bigg) \,,
\end{eqnarray}
where $\Gamma$ denotes any product of Dirac matrices.
Baryons (which are diquarks in this two-color theory) will couple to local
operators of the form
\begin{eqnarray}
{\cal O}_{ud}^{(\Gamma)}(x) &=& u^T(x)(-i\sigma^2)C\Gamma d(x) \,, \\
{\cal O}_{du}^{(\Gamma)}(x) &=& d^T(x)(-i\sigma^2)C\Gamma u(x) \,, \\
{\cal O}_{uu\pm dd}^{(\Gamma)}(x)
    &=& \frac{1}{\sqrt{2}}\bigg(u^T(x)(-i\sigma^2)C\Gamma u(x)
        \pm d^T(x)(-i\sigma^2)C\Gamma d(x)\bigg) \,,
\end{eqnarray}
where the Pauli structure $-i\sigma^2$ acts on color indices while the
charge conjugation operator $C$ acts on Dirac indices.

We extract the meson masses from the two-point correlation functions
\begin{align}
C^{(\Gamma)}_{\overline{u}d}(t_i-t_f)
 & =  \sum_{\vec x_i,\vec x_f} \left\langle {\cal O}_{ud}^{(\Gamma)}(x_f)
{\cal O}_{ud}^{(\Gamma)\dagger}(x_i) \right\rangle.\nonumber\\
 & = \sum_{\vec x_i,\vec x_f} \Tr \Gamma
S_{d\overline{d}}(x_f,x_i)\gamma^0\Gamma^\dagger\gamma^0 S_{u\overline{u}}(x_i,x_f),
\end{align}
where $S_{u\overline{u}}(x,y) = \langle
u(x)\overline{u}(y)\rangle$. The quantities of interest are
pseudoscalar $\Gamma=\gamma_5$, vector $\Gamma=\gamma_k$ ($k=1,2,3$),
and axial vector $\Gamma=\gamma_5\gamma_k$ mesons. As a source vector
we use  $Z_2\times Z_2$ single time slice stochastic sources
\cite{Boyle:2008rh}. 

In addition to the meson spectrum we are interested in two other
quantities, the quark mass $m_{\rm q}$ and the Goldstone boson decay
constant $f_{\rm \Pi}$. We define the quark mass through the Partially Conserved Axial Current (PCAC)
relation: 
\begin{equation}
  m_{\rm q}=\lim_{t \rightarrow \infty}\frac{1}{2}\frac{\partial_t V_{\rm \Pi}}{V_{\rm PP}},
\label{eq:PCAC}
\end{equation}
where
\begin{align}
  V_{\rm \Pi} (t_i-t_f) &= a^3\sum_{x_1,x_2,x_3} \left\langle\overline{u}(t_i)\gamma_0 \gamma_5 d(t_i)\overline{u}(t_f)\gamma_5d(t_f)\right\rangle \ ,\nonumber \\
  V_{\rm PP}(t_i-t_f) &= a^3\sum_{x_1,x_2,x_3} \left\langle\overline{u}(t_i)\gamma_5d(t_i)\overline{u}(t_f)\gamma_5d(t_f)\right\rangle \ .
\end{align}

The Goldstone boson decay constant can be calculated as:
\begin{equation}
  f_{\rm \Pi} = \frac{2 m_{\rm q}}{m_{\rm \Pi}^2}G_{\rm \Pi},
\end{equation}
where $G_{\rm \Pi}$ is obtained from the asymptotic form of $V_{\rm  PP}$ at large $t_i-t_f$:
\begin{equation} 
  V_{\rm PP}(t_i-t_f) \sim -\frac{G_{\rm \Pi}^2}{m_{\rm \Pi}}\exp\left[-m_{\rm \Pi}(t_i-t_f)\right].
\end{equation}

To convert the lattice quantities to physical units, one should determine the lattice spacing for our 
simulations and the appropriate (mass-independent) renormalization constant.
The lattice spacing, in a Technicolor model, is fixed by the requirement that the 
(renormalized) Goldstone boson decay constant has the value of 246 GeV, giving the correct mass to the electroweak 
gauge bosons. For the general composite Higgs scenario the electroweak decay constant becomes $\sin (\theta) \,f_{\Pi}$ with the $\theta$ depending on the specific electroweak embedding. The actual value of $\theta$ depends on the electroweak quantum corrections, the top corrections as well as the effects of other possible sources of explicit breaking of the initial SU($4$) symmetry. The Technicolor limit is recovered for $\theta = \pi/2$ while the composite pGB Higgs case corresponds to small, but non-vanishing, $\theta$. Any other value of the $\theta$ is allowed and corresponds to a combination of these two limits. For the details we refer to \cite{Cacciapaglia:2014uja}. For definitiveness we present the results for $\sin(\theta) = 1$ but we will reinstate the dependence on $\theta$ for the spectrum. 

The relevant renormalization constant for $f_{\rm\Pi}$, commonly denoted in the literature by $Z_a$, 
has not been computed non-perturbatively for our simulations.
In this work we use the perturbative value which has been calculated  in \cite{DelDebbio:2008wb}. 
For fermions in the fundamental representation we have:
\begin{equation}
  Z_a=1-\frac{g_0^2}{16\pi^2}\frac{N^2-1}{2N}15.7\overset{N=2}{=}1-0.2983/\beta.\label{eq:Za}
\end{equation}

\section{The Lattice Results}\label{sec:latresults}

\begin{figure}
\includegraphics[width=0.45\textwidth]{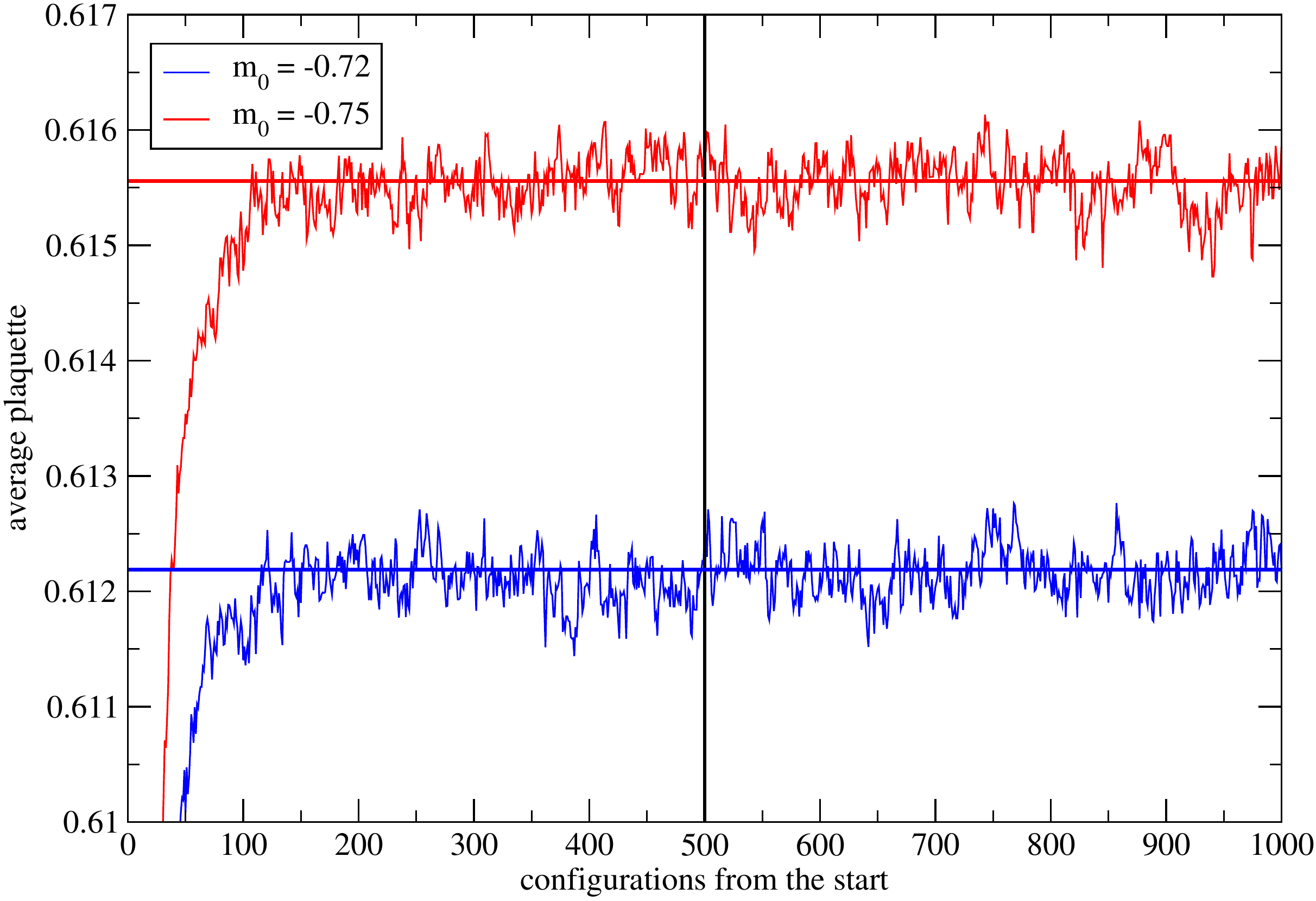}
\includegraphics[width=0.45\textwidth]{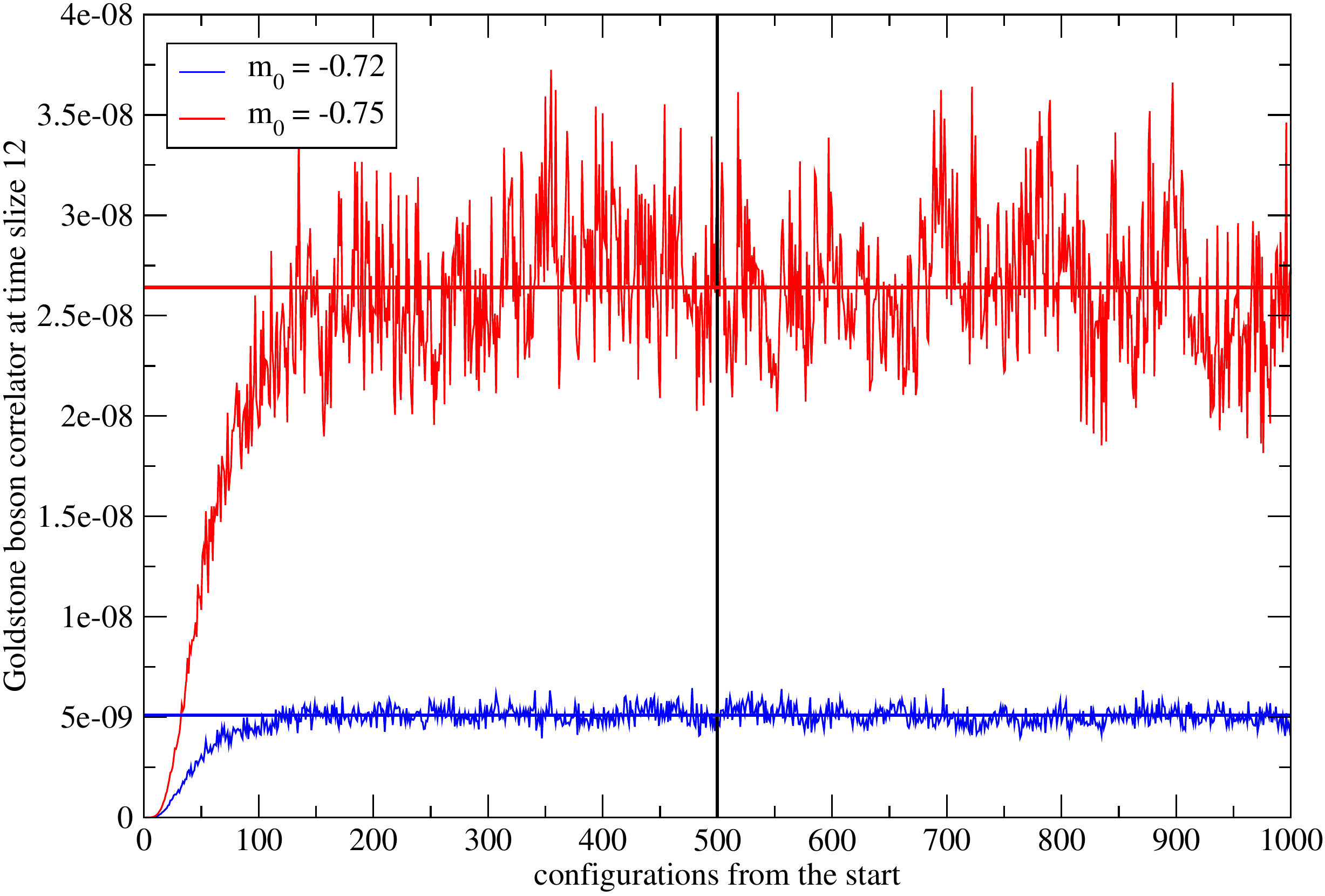}
\caption{Left: The first 1000 measurements of average
  plaquette. Right: The first 1000 measurements of the Goldstone boson
  correlator at time slice 12. The simulations were performed with
  volume $V=32^4$ and coupling $\beta=2.2$. The configurations left of black
  vertical line are discarded as unthermalized.}\label{fig:plaqtherm}
\end{figure}

The lattice simulations used in this work extend the results already published in
Ref.~\cite{Lewis:2011zb}. In particular, we have used larger volumes for the set of parameters closest to the chiral limit. The bare parameters used for our simulations are listed in
Table~\ref{table:sim_param}, where we also report the number of thermalized trajectories, of length one, used in our analysis below. 
Thermalizations are estimated by monitoring the average plaquette expectation value and the value of the two-point correlation function in the pseudoscalar channel at a time separation of twelve time slices. 
These two quantities are shown in Fig.~\ref{fig:plaqtherm} for two representative light quark masses on the finest lattices used in this work.   

\begin{table}
  \begin{tabular}{ccccc}
    \hline
    $\beta$ & Volume & $m_0$ & Therm. & Conf. \\
    \hline
    \hline
    2.0 & $16^3\times32$~~  & -0.85, -0.9, -0.94,
    -0.945, -0.947, -0.949 & 320 & 680 \\
    2.0 & $32^4$ & -0.947 & 500 & 680 \\
    2.2 & $16^3\times32$  & -0.60, -0.65, -0.68,
    -0.70, -0.72, -0.75 & 320 & 680 \\
        2.2 & $24^3\times 32$ &  -0.75 & 500 & $\sim$2000 \\
    2.2 & $32^4$ & -0.72,-0.735, -0.75 & 500 & $\sim$2000 \\
    \hline
  \end{tabular}
  \caption{Parameters used in the simulations. The thermalization
    column refers to the number of discarded initial configurations
    and the configuration column refers to the number of independent
    configurations used in measurements.  \label{table:sim_param}}
\end{table}

\begin{table}
  \begin{tabular}{cccccccc}
    \hline
    $\beta$ & Volume & $m_0$ & $m_q$ & $m_{\rm \Pi}$ & $m_{\rm \rho}$ &
    $m_{\rm A}$ &$f_{\rm \Pi}$
     \\
    \hline\hline
2.0 & $32\times16^3$ & -0.85 & 0.1919(6) & 0.9163(18) & 1.008(20) & 1.64(3) & 0.1544(4)\\
2.0 & $32\times16^3$ & -0.9 & 0.1134(6) & 0.708(3) & 0.821(3) & 1.47(3) & 0.1248(5)\\
2.0 & $32\times16^3$ & -0.94 & 0.0476(8) & 0.451(7) & 0.636(5) & 1.15(4) & 0.089(9)\\
2.0 & $32\times16^3$ & -0.945 & 0.038(7) & 0.407(5) & 0.57(6) & 1.08(3) & 0.0799(7)\\
2.0 & $32\times16^3$ & -0.947 & 0.0327(7) & 0.377(6) & 0.546(7) & 1.02(4) & 0.0754(8)\\
2.0 & $32\times16^3$ & -0.949 & 0.0307(8) & 0.374(6) & 0.546(9) & 1.02(3) & 0.075(12)\\
2.0 & $32^4$ & 0.947 & 0.0309(3) & 0.3739(14) & 0.536(4) & 1.0(5) & 0.0766(6)\\
2.2 & $32\times16^3$ & -0.6 & 0.2296(7) & 0.886(3) & 0.93(3) & 1.371(12) & 0.1119(5)\\
2.2 & $32\times16^3$ & -0.65 & 0.1637(7) & 0.755(3) & 0.792(3) & 1.229(9) & 0.0992(5)\\
2.2 & $32\times16^3$ & -0.68 & 0.1205(7) & 0.612(4) & 0.671(5) & 1.068(17) & 0.0868(5)\\
2.2 & $32\times16^3$ & -0.7 & 0.0968(7) & 0.548(5) & 0.615(6) & 1.018(12) & 0.0793(5)\\
2.2 & $32\times16^3$ & -0.72 & 0.0686(6) & 0.455(5) & 0.531(5) & 0.884(19) & 0.0684(5)\\
2.2 & $32\times16^3$ & -0.75 & 0.0264(8) & 0.324(8) & 0.445(9) & 0.76(3) & 0.0405(12)\\
2.2 & $32\times24^3$ & -0.75 & 0.024(5) & 0.258(4) & 0.359(8) & 0.62(4) & 0.0433(9)\\
2.2 & $32^4$ & -0.72 & 0.0661(5) & 0.4475(8) & 0.522(14) & 0.81(3) & 0.0664(5)\\
2.2 & $32^4$ & -0.735 & 0.0456(3) & 0.3612(18) & 0.446(4) & 0.75(3) & 0.0568(5)\\
2.2 & $32^4$ & -0.75 & 0.0257(5) & 0.2649(16) & 0.363(5) & 0.59(6) & 0.0457(7)\\
\hline
  \end{tabular}
  \caption{The values obtained in simulations for PCAC-quark mass, Goldstone
    boson mass, vector meson mass, axial vector meson mass, and
    Goldstone boson decay constant as function of $\beta$, volume and
    bare quark mass.} 
\end{table}

All the ensembles of gauge configurations were created using the GPU version of the HiRep code
\cite{DelDebbio:2008zf}. The lattice action used is the plaquette-action SU(2)
gauge theory with two flavors ($u$ and $d$) of mass-degenerate Wilson
fermions.  The Hybrid Monte Carlo trajectory length was chosen to be one. The autocorrelation
times for plaquette expectation values and meson correlators were estimated to be about 10. 
The errors for all quantities extracted in this work were obtained using a bootstrap procedure.

\begin{figure}
  \includegraphics[width=0.9\textwidth]{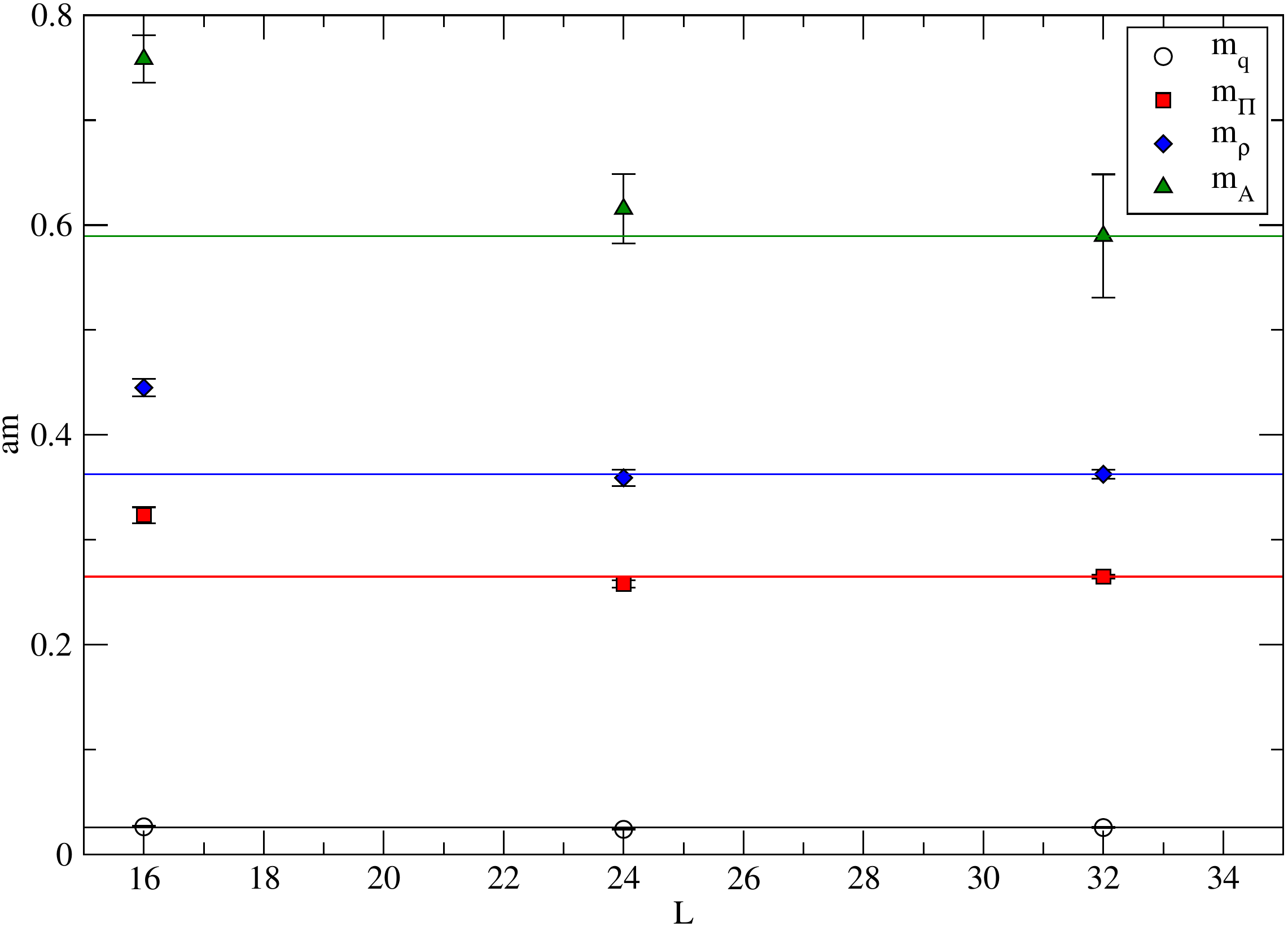}
  \caption{The PCAC quark mass, Goldstone boson, vector meson, and axial vector
    meson mass as a function of lattice size $L$. On the most chiral
    point $m_0=-0.75$ and $\beta=2.2$. The measurements on two larger
    lattices are inside statistical errors.\label{fig:finitevol}}
\end{figure}

\begin{figure}
  \includegraphics[width=0.9\textwidth]{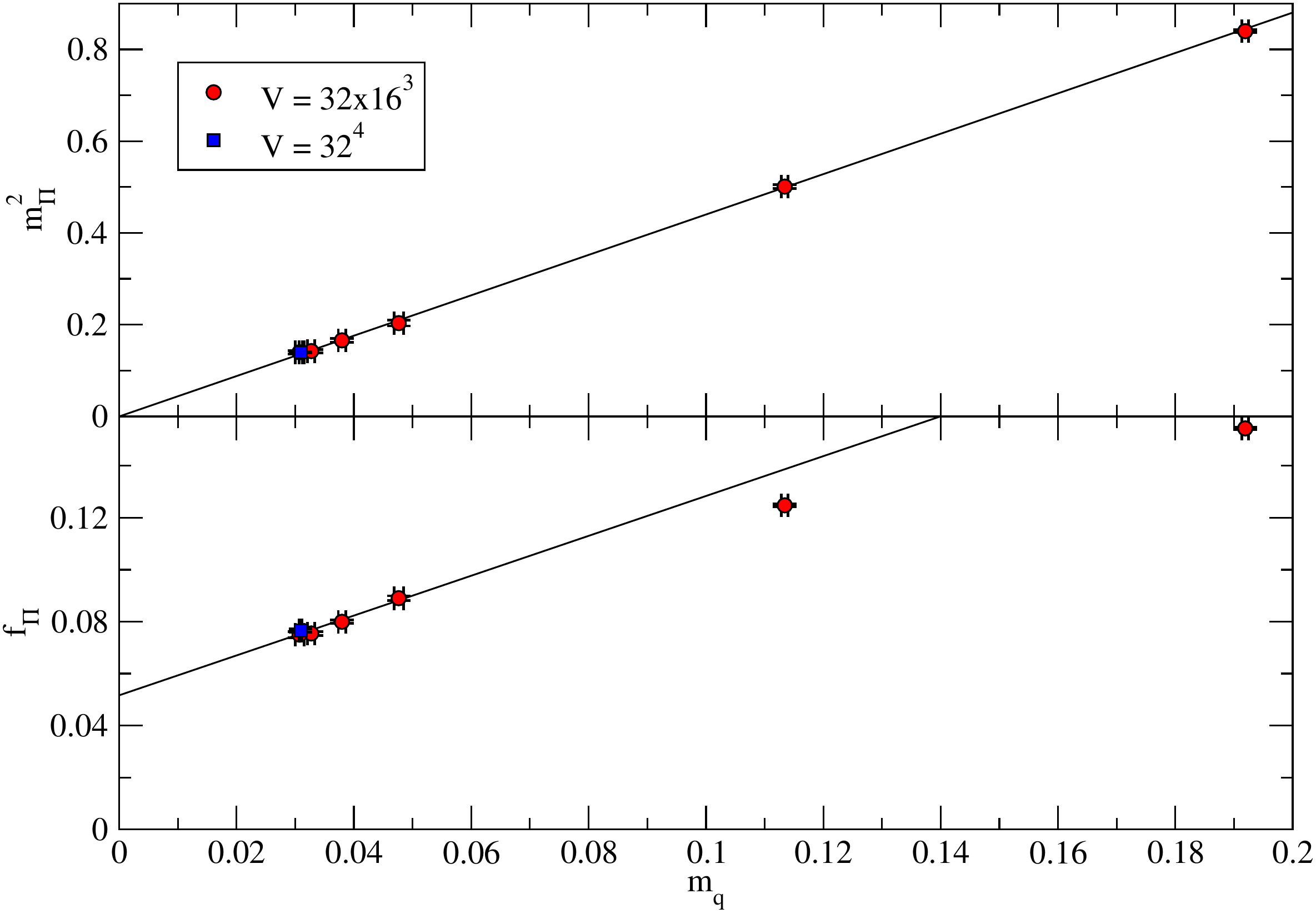}
  \caption{The Goldstone boson mass squared and its decay constant as a function of the quark
    mass for $\beta=2.0$. Extrapolations to the chiral limit, as discussed in the text, 
    are also shown.\label{fig:chiral20}}
\end{figure}

\begin{figure}
  \includegraphics[width=0.9\textwidth]{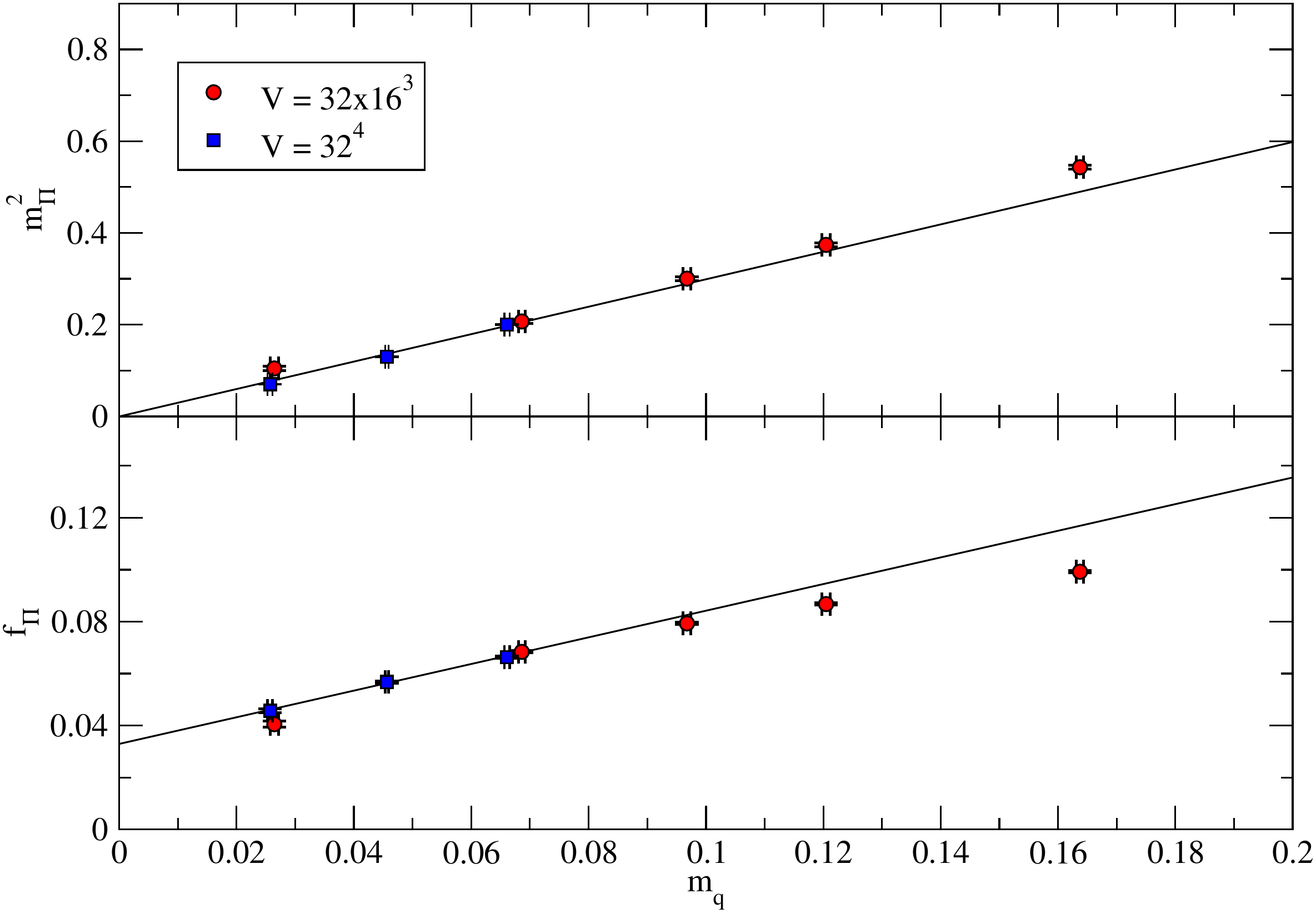}
  \caption{The Goldstone boson mass squared and its decay constant as a function of the quark
    mass for $\beta=2.2$. Extrapolations to the chiral limit, as discussed in the text, 
    are also shown.\label{fig:chiral22}}
\end{figure}

\begin{figure}
  \includegraphics[width=0.9\textwidth]{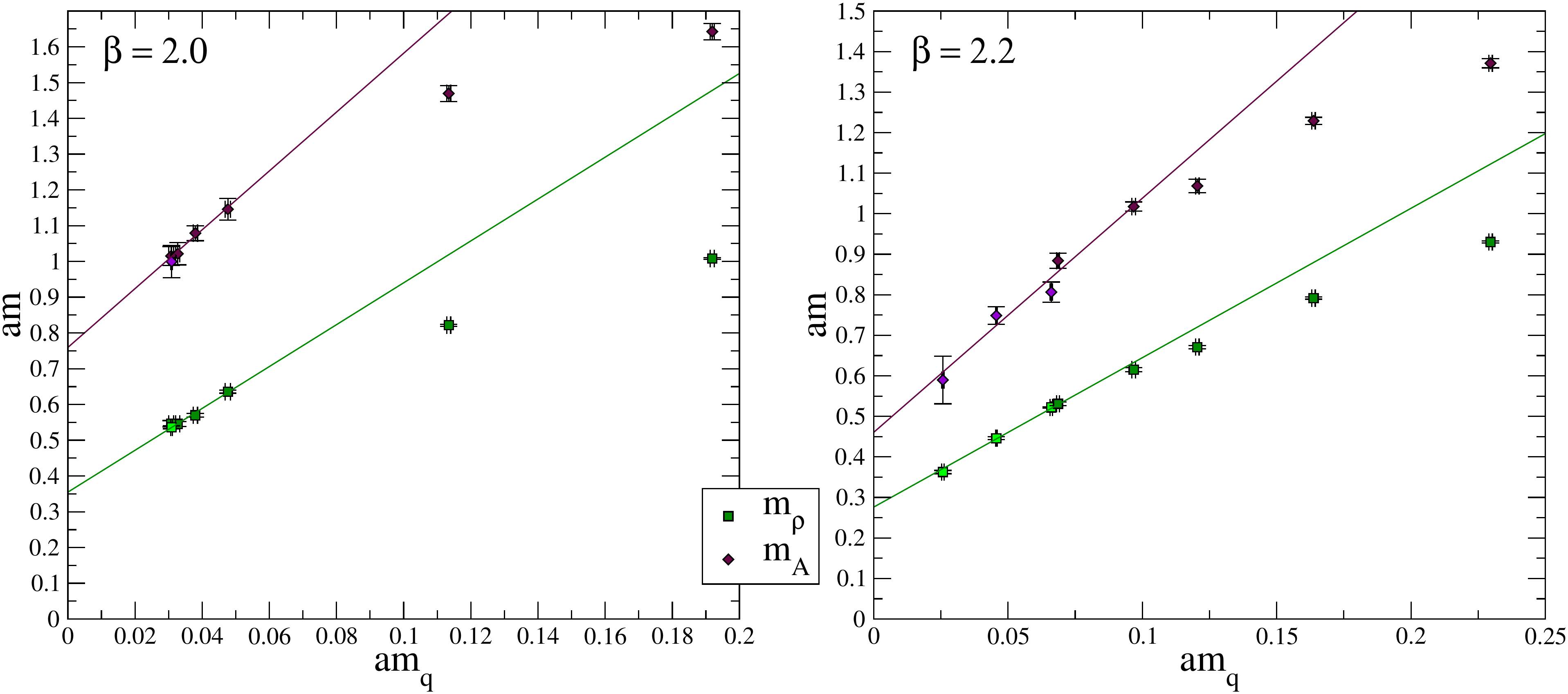}
  \caption{Vector and axial vector meson as a function of PCAC quark
    mass with two different lattice spacings. A linear extrapolation to the
    chiral limit works well with $m_q<0.1$. \label{fig:spectrum}}
\end{figure}

\begin{figure}
  \includegraphics[width=0.9\textwidth]{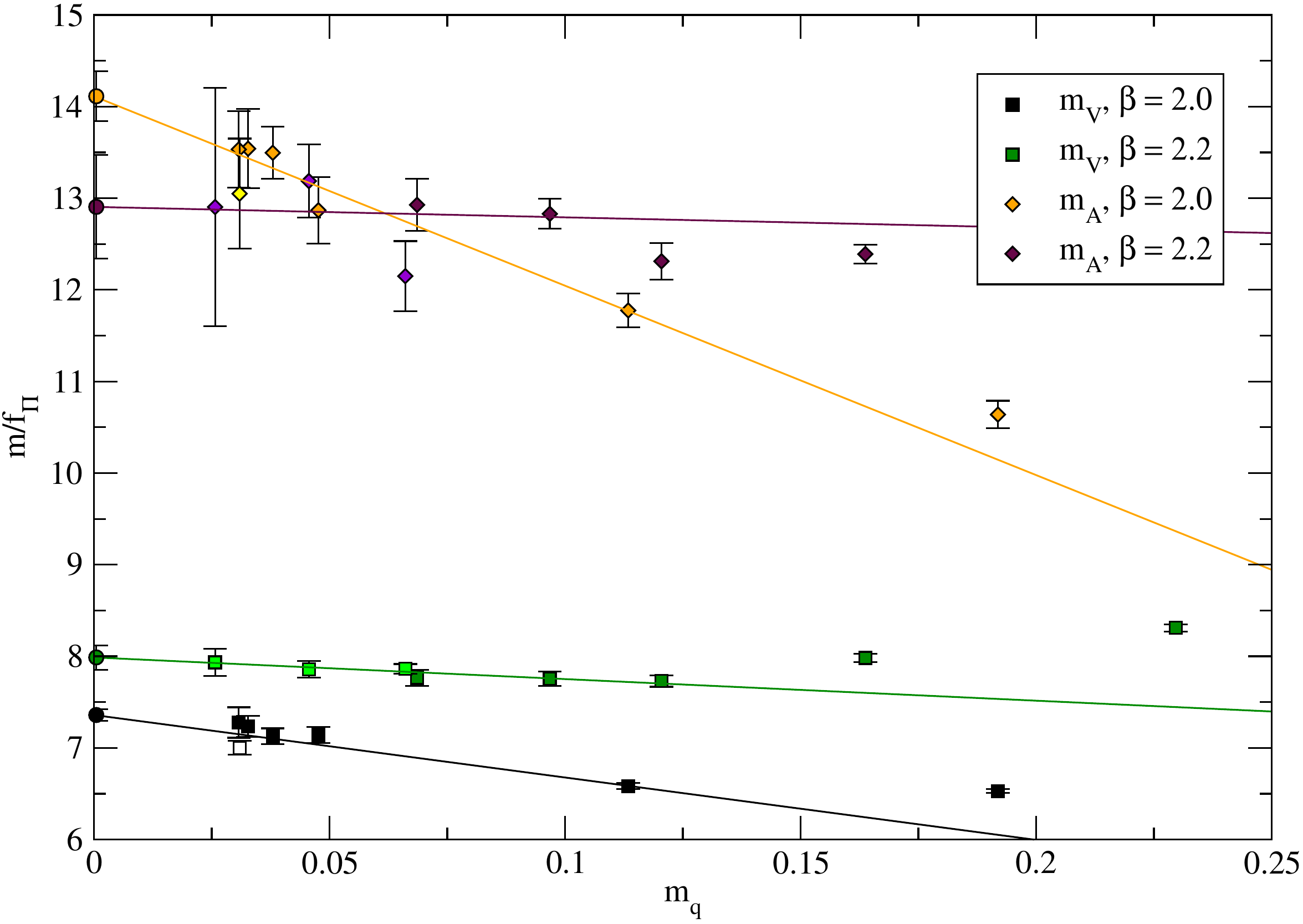}
  \caption{The vector meson and axial vector meson masses in physical units. The chiral extrapolations have been performed using a linear fit to the points where $m_q<0.12$. \label{fig:spectrum_phys}}
\end{figure}

\begin{figure}
  \includegraphics[width=0.9\textwidth]{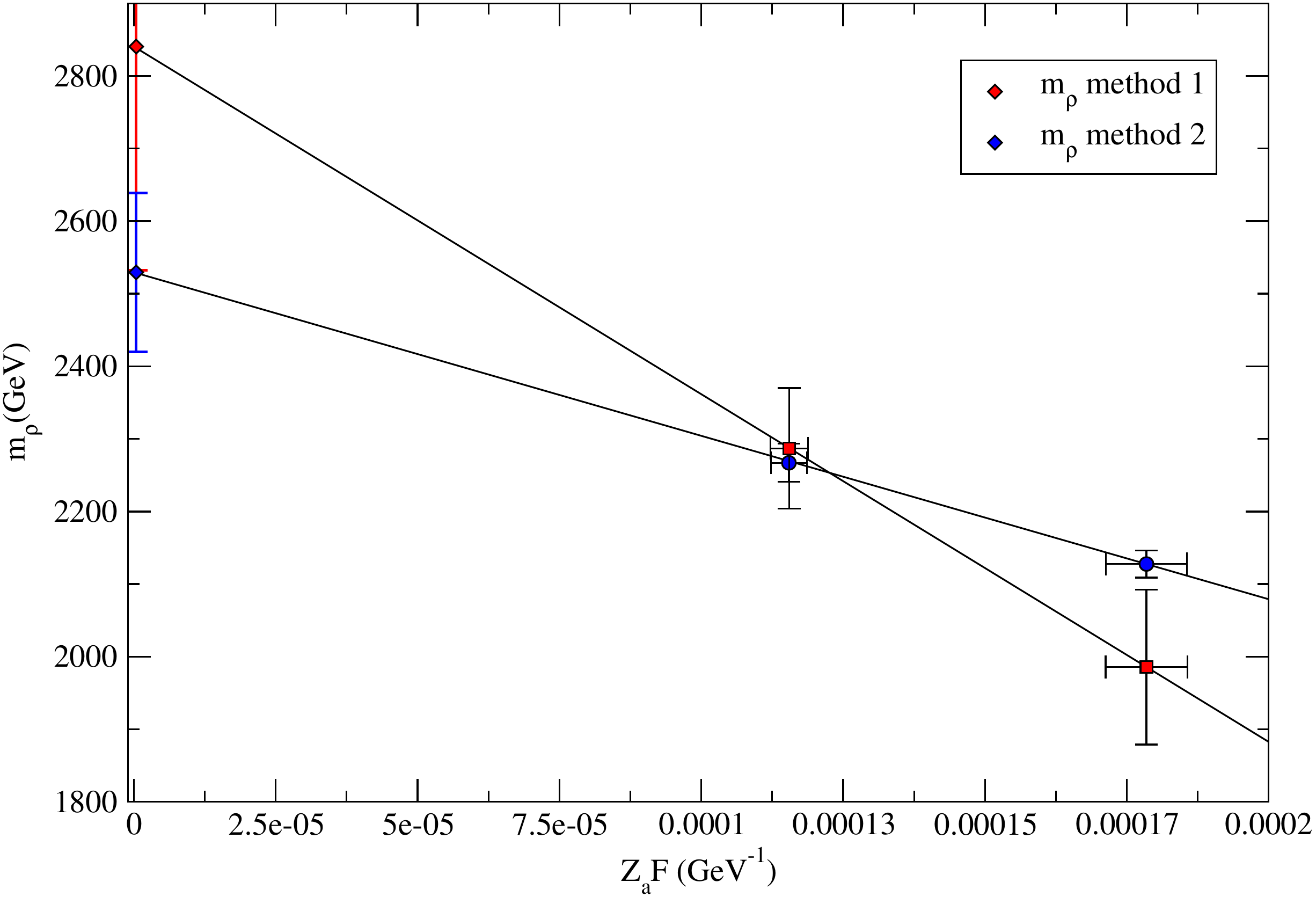}
  \caption{A continuum extrapolation of the vector
    meson mass. \label{fig:mv_cont}}
\end{figure}

In a previous work by some of the authors~\cite{Lewis:2011zb}, a first estimate of the Goldstone spectrum was already obtained.
However large finite volume effects were observed -- see Fig.~5(b) of \cite{Lewis:2011zb} -- for the lightest value of the quark mass on the finest lattice used in that work, corresponding to the bare parameter couplings $(\beta,m_0)=(2.2,-0.75)$.

Here we perform a more systematic analysis to control finite volume effects, using simulations on three different lattice volumes $V=32\times16^3$, $32\times24^3$, and $32^4$ at the lightest quark mass on the finest lattice. 
At this quark mass, the results for the smallest volume $32\times16^3$ suffer clearly from finite volume effects, whereas observables measured on the two largest lattices agree within statistical errors, as shown in Fig.~\ref{fig:finitevol}. 
From the size of the statistical errors on the two largest lattices, we estimate 
that the residual finite volume effects at our lightest quark mass are below 2\% 
for  $m_{\rm \Pi}$, $f_{\rm \Pi}$ and $m_\rho$ and below 10\% for $m_A$.

To confirm spontaneous chiral symmetry breaking one should, in principle, reach the chiral regime of the theory, pushing the quark masses light enough that chiral perturbation theory ($\chi$PT), or the appropriate lattice extension of it, could be used, while keeping under control all other systematic sources of error which are present on the lattice. It is well known, by studies of QCD, that this chiral regime is extremely difficult to reach as lattice artifacts and residual finite volume effects tend to make the predictions of $\chi$PT difficult to test. 

Keeping this in mind, we analyze our data for signs of spontaneous chiral symmetry breaking and check how well predictions from $\chi$PT fit the measured Goldstone spectrum.
The Goldstone boson mass and decay constant are studied as a function of quark mass, defined through the PCAC relation and compared with the expectations from (continuum) chiral perturbation theory at next to leading order (NLO):
\begin{equation}
  \frac{m^2_{\rm \Pi}}{m_{\rm q}} = 2B\left[1 + C x\log x + D x +
  \mathcal{O}(m_{\rm q}^2)\right]\,\,,
  \label{eq:chiralgb}
\end{equation}
and 
\begin{equation}
  f_{\rm \Pi} = F\left[1 + C' x\log x + D' x +
  \mathcal{O}(m_{\rm q}^2)\right]\,\,,
  \label{eq:chiralfpi}
\end{equation}
where $B$, $F$, $D$ and $D'$ are (unknown) low-energy constants of the
theory, $x\equiv\frac{2B m_{\rm q}}{16\pi^2 F^2}$ and $C$ and $C'$ are
known constants. For our theory $C=-\frac12-\frac{1}{2N_f}=-\frac34$
and $C'=\frac12N_f = 1$.
Additional terms in the chiral expansion can also be computed for the NNLO approximations.
The relevant expressions can be found in \cite{Bijnens:2009qm}, from which the 
quoted values for $C$ and $C'$ were taken.

Correction to the continuum chiral expansion arise due to the lattice discretisation at non-zero lattice spacing.
For Wilson fermions at NLO the functional form of Eqs.~(\ref{eq:chiralgb}) and (\ref{eq:chiralfpi})
remains unchanged, but the coefficients, and in particular $C$ and $C'$, 
depend on the lattice spacing $a$. In the limit $a\rightarrow 0$ one should recover the continuum values
for $C$ and $C'$, however for a fixed lattice spacing $C$ and $C'$ are two additional free 
parameters of the expansion.

We show in Figs.~\ref{fig:chiral20} and \ref{fig:chiral22} our results for $m^2_{\rm \Pi}$ 
and $f_{\rm \Pi}$ for the two values of the lattice spacing used in
this work. It is possible to use NLO Wilson chiral perturbation theory
to fit our data at small quark masses for both $m^2_{\rm\Pi}$ and
$f_{\rm \Pi}$. We report in Table~\ref{table:xptfit} the quark mass
ranges, $\chi^2$ and the coefficients $B$ and $F$ for the fits for the
two different values of the lattice spacing used in this work. The
relative errors on the fitting parameters are large especially for
the coefficients $C$ and $C'$  of the $x\log x$ terms of
the chiral expansion which suffer from very large uncertainties $\sim
100\%$,  and are thus compatible with zero. This can be explained by our data not being yet in a regime where
the $x\log x$ terms can be clearly distinguished from the polynomial terms in the expansion, even at the lightest quark 
masses available.


\begin{table}
  \begin{tabular}{cccccc}
    \hline
     & $\beta$ & range & ~$\chi^2$/dof~  & ~dof~ & ~LO coefficient  \\
    \hline\hline
    $m^2_{\rm \Pi}~$ & ~2.0~ & ~$m_q<0.05$~ & 0.48 & 2 & $B=4.2(2.3)$ \\
    $f_{\rm \Pi}$ & 2.0 & $m_q<0.05$ &  2.28 & 2 & $F=0.13(5)$ \\
    $m^2_{\rm \Pi}$ & 2.2 & $m_q<0.07$ & 0.32 & 1 & $B=1.34(16)$\\
    $f_{\rm \Pi}$ & 2.2 & $m_q<0.07$ & 2.01  & 1 & $F=0.028(6)$ \\
    \hline
  \end{tabular}
  \caption{NLO Wilson $\chi$PT fits to our data for 
  $m^2_{\rm\Pi}$ and $f_{\rm\Pi}$. All the fits are acceptable in the quoted quark mass range. The 
  last column ``LO coefficient" refers to the coefficients $B$ and $F$ of the chiral expansion for the 
  Goldstone boson mass squared and for its decay constant.\label{table:xptfit}}
\end{table}

Given that the chiral logs are subdominant, a simple polynomial fit 
to the data is expected to be an adequate description of $m^2_{\rm \Pi}$ and  $f_{\rm \Pi}$.
We therefore fit our data setting $C=C'=0$. The values for $B$ and $F$ thus obtained are
given in Table~\ref{table:gbdecay}.  The central values of $B$ and $F$ are 
compatible, within statistical errors, with the ones obtained using NLO Wilson chiral perturbation
theory.  In the analysis below we use these values of $B$ and $F$ as our best estimate, and consider 
the errors from different fitting procedures as systematic errors. In the last column of Table~\ref{table:gbdecay},
we use the perturbative value of $Z_a$ from Eq.~(\ref{eq:Za}) to obtain the renormalized $F$. 

We also note that our data are not well described by NLO or NNLO continuum chiral
perturbation theory, i.e. when the coefficients $C$ and $C'$ of the logarithmic terms are fixed. 
In this case we can perform a simultaneous fit of both $m^2_{\rm \Pi}$ and  $f_{\rm 
\Pi}$, and the fit is thus much more constrained. The resulting $\chi^2/{\rm dof}\sim 100$ shows that
this is not a good description of our data. 

\begin{table}
  \begin{tabular}{crrrr}
    \hline
    $\beta$ & \multicolumn{1}{c}{$B$} & \multicolumn{1}{c}{$F$}& \multicolumn{1}{c}{$Z_a$} &\multicolumn{1}{c}{$Z_a F$}  \\
    \hline\hline
    2.0 & 2.52(12) & 0.052(3) & 0.85 & 0.0439(18) \\
    2.2 & 1.26(03) & 0.033(1) & 0.86 & 0.0285(08) \\
    \hline
  \end{tabular}
  \caption{The fitted values for the coefficients $B$ and $F$ of the chiral expansion. The functional form
  used is a polynomial in the quark mass, as explained in the text. The corresponding $\chi^2/{\rm  dof}$
  are 0.51 and 3.4 for $\beta=2.0$, and 0.28 and 1.5 for $\beta=2.2$ for $m^2_{\rm\Pi}$ and 
  $f_{\rm\Pi}$ respectively. In the last column we report the value of the renormalized $F$ obtained 
   using the perturbative value of $Z_a$. \label{table:gbdecay}}
\end{table}




The values of vector and axial vector meson masses, as measured from
our emsemble of configurations, are plotted in
Fig.~\ref{fig:spectrum}. A linear function 
represents the data well at small quark masses $m_{\rm q}<0.1$. 


Given that $f_{\rm\Pi}$, $m_\rho$ and $m_A$ are well described by a linear 
function at small quark masses, one can expect that the two ratios $m_\rho/f_{\rm\Pi}$ and
$m_A/f_{\rm\Pi}$ are also linear functions of the quark mass close to the chiral 
limit. These ratios are shown in Fig.~\ref{fig:spectrum_phys}, together with a 
linear extrapolation to zero quark mass. We will refer to this method of chiral 
extrapolation as ``method 2'', whereas the first method will be named ``method 
1'' in the following.

We use these two different methods as a crosscheck of the chiral extrapolation and to try to quantify 
the systematic errors due to the choice of extrapolation function. We compare in Table~\ref{table:chiral} 
the results in lattice units for the chiral extrapolations. The methods are clearly consistent with each 
other and method 2 leads to an overall smaller statistical error.

Combining the data from the two lattice spacings available in this study, we can 
perform a first, crude continuum extrapolation for the masses of the vector and axial vector mesons. 
As explained above, the lattice spacing is fixed by the requirement that the value of the renormalized Goldstone 
decay constant satisfies $f_\Pi \sin(\theta) = 246~{\rm GeV}$, as required to give the correct masses to the electroweak gauge bosons. For concreteness here we assume $sin(\theta) = 1$, but the dependence on $\theta$ can easily be reinstated when required as done below. The results of the linear extrapolations of $m_\rho/(Z_a F)$ and $m_A/(Z_a F)$ to the 
continuum limit are reported in  Table~\ref{table:cont3}.

Our continuum extrapolation is subject to two major sources of systematic
errors. First our simulations are performed only at two lattice
spacings, and therefore we do not have a good measure of how well our
linear extrapolation describes the data. To take this into account, as
a systematic error we quote, quite conservatively, the difference
between the value of the continuum extrapolated value and the data 
point of the finer lattice.  The second systematic error stems
from the renormalization constant $Z_a$ which we do not measure
non-perturbatively. As a systematic error we then use the difference
between the perturbative value of $Z_a=Z_{\rm a}^{\rm pert}$ and
$Z_{\rm a}=1$. In Table~\ref{table:cont3} we list the continuum
extrapolated values of vector and axial vector mesons for both methods
1 and 2 described above for chiral extrapolation. 
The vector case is plotted in Fig.~\ref{fig:mv_cont}.

The results produced by both methods are comparable and well inside
each other's error bars. As final results for the meson masses we
quote the one obtained by method 2. Square summing the errors, the
vector meson reads $m_\rho \sin (\theta)=2.5\pm 0.5~{\rm TeV}$ and the axial vector
meson $m_A \sin (\theta)=3.3  \pm 0.7~{\rm TeV}$ where we have reinstated the dependence on the angle $\theta$ defining the specific electroweak embedding. 

\begin{table}
  \begin{tabular}{crrrr}
    \hline
     & \multicolumn{1}{c}{$\beta$} &  \multicolumn{1}{c}{Method 1} & \multicolumn{1}{c}{Method 2} \\
    \hline\hline
    $m_\rho/(Z_a F)$ & 2.0 & 8.1(5) & 8.65(8) \\
    $m_\rho/(Z_a F)$ & 2.2 & 9.3(4) & 9.22(11) \\
    $m_A/(Z_a F)$ & 2.0 & 18(2)  & 16.6(4)  \\
    $m_A/(Z_a F)$ & 2.2 & 17(3)  & 15.5(6) \\
    \hline
  \end{tabular}
  \caption{The chiral extrapolated values for the vector
    and axial vector mesons in units of the renormalized Goldstone boson decay constant. 
    Only the statistical error is reported here.\label{table:chiral}}
\end{table}

\begin{table}
  \begin{tabular}{crrrr}
    \hline
     & \multicolumn{1}{c}{$\beta$} &  \multicolumn{1}{c}{Method 1 (GeV)} & \multicolumn{1}{c}{Method 2 (GeV)} \\
    \hline\hline
    $m_\rho$ & $\infty$\; & 2840(330)(560)(360) & 2520(100)(240)(310)  \\
    $m_A$ & $\infty$\; & 4000(1800)(200)(430)  & 3300(400)(510)(340) \\    
    \hline
  \end{tabular}
  \caption{The continuum extrapolated values for the vector
    and axial vector mesons. The conversion to physical units is done requiring $Z_a F=a\cdot 246~{\rm GeV}$.
    The first error  is statistical, the second one is the systematic
    from a linear continuum extrapolation with only two data 
    points, and the third one comes from the uncertainty on $Z_a$.\label{table:cont3}}
\end{table}

\section{Conclusions\label{sec:conclusions}}
The SU(2)-gauge theory with two fundamental fermions unifies both
Technicolor and composite pGB Higgs models of electroweak symmetry
breaking. In this work, we have calculated the masses of the two
lightest non-singlet mesons using the Goldstone boson decay 
constant to set the scale.  We performed the calculations with
two  different lattice spacings. With conservative error 
estimates the mass of the lightest vector meson $m_\rho  \sin (\theta)$ is $
2.5\pm 0.5 {\rm TeV}$. This value is clearly above one TeV  and outside the current
exclusion limits set by the LHC \cite{Aad:2012hf}.

To increase the precision of our results for the spectrum, at least one additional
lattice spacing is required alongside a nonperturbative determination of the renormalization constant
$Z_a$.  This would require a significant increase of computational resources. Furthermore we are eager to 
investigate the scalar sector and the vector decay constants.  

\acknowledgments

This work was supported by the Danish National Research Foundation
DNRF:90 grant, by a Lundbeck Foundation Fellowship grant, and by NSERC
of Canada. The computing facilities were provided by the Danish Centre for 
Scientific Computing and Canada's Shared Hierarchical Academic
Research Computing Network (SHARCNET: http://www.sharcnet.ca).

\end{document}